\documentclass[aps,prb,twocolumn,superscriptaddress,preprintnumbers,amsmath,amssymb]{revtex4-2}
\usepackage{graphicx}
\usepackage{dcolumn}
\usepackage{bm}
\usepackage{epstopdf}
\usepackage{xcolor}
 \usepackage{multirow}
\usepackage{makecell}
\usepackage{chngcntr}
\usepackage[hyperfigures=true,bookmarksnumbered=true,bookmarksopen=true,bookmarks=true,linkcolor=blue,urlcolor=blue,citecolor=blue,colorlinks=true,pdfhighlight=/O,pdfstartview={XYZ null null 1}]{hyperref}

\begin{document}
\title{Pressure-Driven Transitions in La$_2$CoTiO$_6$: Antiferromagnetic Insulator to Nonmagnetic Metal via Antiferromagnetic Metal in a Double Perovskite Oxide}

\author{Sromona Nandi}
\affiliation{Department of Physics, Indian Institute of Technology Tirupati, Tirupati 517619, AP, India}

\author{Subhadip Pradhan}
\affiliation{School of Physical Sciences, National Institute of Science Education and Research, An OCC of Homi Bhabha National Institute, Jatni 752050, India}

\author{Ashis K. Nandy}
\email[]{aknandy@niser.ac.in}
\affiliation{School of Physical Sciences, National Institute of Science Education and Research, An OCC of Homi Bhabha National Institute, Jatni 752050, India}

\author{Rudra Sekhar Manna}
\email[]{rudra.manna@iittp.ac.in}
\affiliation{Department of Physics, Indian Institute of Technology Tirupati, Tirupati 517619, AP, India}

\date{\today}


\begin{abstract}

In double perovskite oxides (A$_2$BB$^\prime$O$_6$), magnetism often arises from diluted magnetic lattices, created by combining a perovskite structure with localized 3$d$ magnetic elements (B) alongside another perovskite lattice containing nearly nonmagnetic delocalized 4$d$/$5d$ elements (B$^\prime$). Alternatively, the magnetic lattice can consist entirely of 3$d$ elements, with one being completely nonmagnetic with $d^0$ state. La$_2$CoTiO$_6$ (LCTO), a representative double perovskite oxide, contains Ti in a nonmagnetic state with a $d^0$ electron configuration due to its $4^+$ oxidation state. Experimental evidence shows that LCTO possesses a monoclinic structure (space group $P2_1/n$) and behaves as an antiferromagnet with a N\'{e}el temperature of 14.6 K. Through first-principle electronic structure calculations, we uncover that adjusting external hydrostatic pressure induces a sequence of phase transitions: from antiferromagnetic insulator (AFM-I) to antiferromagnetic metal (AFM-M), and ultimately to itinerant nonmagnetic metal (NM-M). The transition from AFM-I to AFM-M at $\sim$ 42 GPa pressure coincides with a shift in spin states, moving from a high-spin (HS) state to a low-spin (LS) state, while Co retains a $d^7$ configuration. Distortion within the monoclinic structure under pressure plays a pivotal role in the spin-state transition. At the AFM-I to AFM-M transition, we observe a sharp decrease in the ratio of the octahedral volumes occupied by Co and Ti. Such change in ratio is linked to variations in octahedral volumes, akin to a breathing mode distortion. We explore the impact of the breathing mode distortion by examining a highly symmetric theoretical structure (space-group $I4/mmm$), achieved by optimizing the structure with all $\angle${Co-O-Ti} angles set to 180$^{\circ}$. Remarkably, the LS state in the LCTO theoretical structure persists under ambient pressure conditions, underscoring the unique role of breathing mode distortion in the monoclinic phase, facilitating the HS to LS transition under pressure. In the LS state, LCTO displays metallic behavior, even with a substantial local correlation (Hubbard $U$) on Co, as large as $U$ = 6 eV. The spin-state transition is further elucidated through an energy level diagram, illustrating a significant modification in the crystal field splitting between Co-$t_{2g}$ and -$e_g$ levels, driven by the robust hybridization of Co-$d$ and O-$p$ orbitals. Finally, with a further increase in pressure, the system attains the NM-M phase at $\sim$ 130 GPa, leading to the complete suppression of the magnetic moment on Co.

\end{abstract}

\maketitle

\section{Introduction}
\label{introduction}

Double perovskites (DPs) have been attracted significant attention in the last few decades due to their unique electronic and magnetic properties owing to the interplay between charge, spin, orbital and lattice degrees of freedoms~\cite{Khomskii1997,Murakami1998,Dagotto2008,Arkadeb2022}. This has general chemical formula A$_2$BB$^\prime$O$_6$, exhibiting a wide range of compositions due to the natural ability to form perovskite structures in double order which tolerates the altering of multiple transition-metal (TM) cations B and B$^\prime$ of varying sizes and electronic configurations. These TMs are surrounded by oxygen anions forming octahedral units and the A site cation refers to the various alkaline-/rare-earth elements~\cite{Mitchell2002}. Because of their numerous physical phenomena such as half-metallicity, above room temperature ferro-/ferrimagnetism, thermoelectric, magneto-dielectrics, magneto-optic and multiferroic properties, semiconducting behavior $etc$, DPs have received a lot of attention for technological applications, particularly in spintronics~\cite{Iliev2008,Nakagawa1968,Fiebig2002,Kobayashi1998,Kobayashi1999}. Among them, Sr$_2$FeMoO$_6$~\cite{Sarma2000,Sarma2001}, Sr$_2$FeReO$_6$~\cite{Longo1961,Pandey2022} etc. are reported as ferrimagnetic (FiM) half-metallic DPs. In contrast to metallic FM DPs, there are antiferromagnetic DPs, $e.g.$, Sr$_2$FeWO$_6$~\cite{Kawanaka1999} having an insulating ground state can be transformed to FiM metal if one dopes Mo at the W site, $i.e.$, Sr$_2$FeMo$_x$W$_{1-x}$O$_6$ with $x \ge 0.3$~\cite{Nakagawa1969,Ray2001}. In addition to that, a list of DPs exhibit Mott insulating  behavior due to the delicate balance among spin-orbit coupling (SOC), on-site Coulomb ($U$) repulsion, and the crystal field effect on the 4$d$/5$d$ TM sites~\cite{Feng2016,Cavichini2018,Ghimire2016,Chen2018,Narayana2010}. In fact, there are reports for the pressure induced metal-insulator transition (MIT) for perovskite materials~\cite{Cai2007,Qiu2017,Bennett2022,Cheng2010} which, however, is hardly explored in case of DPs. MIT transition is also reported for DPs with partial or full chemical substitution on the A and/or TM sites~\cite{Poddar2004,Nazir2021,Krockenberger2007}. In addition, there are a few FM/FiM DPs where the spin-state transition is accompanied by a transition from an insulator to nonmagnetic metal via a half-metallic state~\cite{Lv2012,Zhao2015} and half-metallic to nonmagnetic metal via a semiconducting state~\cite{Qian2012}. However, so far, best of our knowledge, the MIT in AFM DPs associated with the spin-state transition and finally, a nonmagnetic metallic state as function of applied pressure is neither observed in experiments nor theoretically prescribed.

One interesting class of double perovskite with $d^0$ electronic configuration at B$^\prime$ site, $e.g.$, La$_2$CoTiO$_6$ (LCTO), has moderately been studied in experiments~\cite{Rodriguez2002,Holman2007,Yuste2011}. This is rather a simple system with Ti in $4^+$ oxidation state suggesting $d^0$ electronic configuration. The compound shows an antiferromagnetic ground state with low N\'eel temperature ($T_{\mathrm{N}}$ = 14.6 K). This is due to the fact that magnetic lattice is formed by the Co atoms only and hence, it is diluted by the presence of nonmagnetic Ti atom in LCTO. The nearest-neighbor distance between two Co is roughly $\sqrt{2}$ times large compared to that in perovskite oxides~\cite{Middey2012,Nandy2013}. In general, therefore, relatively weak magnetic exchange coupling between Co magnetic moments is expected. From simple charge balance calculation, Co is expected to show $d^7$ electronic configuration due to the Co$^{2+}$ oxidation state. Neutron diffraction reveals the long-range antiferromagentic order is due to the arrangement of Co spins~\cite{Holman2007} with a propagation vector [k = 1/2, 0,1/2]~\cite{Rodriguez2002}. There are some recent studies regarding the cobaltate-titanate oxide heterostructures, hinting about the orbital polarization of Co$^{2+}$ state as a function of electronic correlation and by breaking of electronic symmetry with the formation of a superlattice. Experimentally such a superlattice can be grown as an epitaxial thin film on a substrate~\cite{Lee2019,Lee2021}. Such $d$ electron configuration ($d^7$) may result in altering Co magnetic moment from its HS value to LS value or vice-versa due to the change in octahedral geometry. In contrast to the Co$^{2+}$ ($d^7$) in LCTO, a recent study~\cite{Takegami2023} considering a perovskite, LaCoO$_3$ with Co$^{3+}$ ($d^6$), has shown HS to LS transition associated with the changes in the geometry of CoO$_6$ octahedra as a function of temperatures.

In this work, within combined experimental and theoretical approaches, we characterize the structural and magnetic behavior of LCTO at ambient pressure. The pressure effects have been then studied within first-principle electronic structure calculations in order to explore the electronic and the associated magnetic properties of LCTO. At ambient pressure, the prepared LCTO has a monoclinic structure with space-group $P2_1/n$ as determined from the X-Ray diffraction (XRD) and shows an AFM ground state from the Curie-Weiss analysis. The magnetic ground state is consistent with the electronic structure calculations. The insulating nature of having band gap 1.01 eV estimated from the total density of states (DOS) is consistent with the reported resistivity measurement~\cite{Holman2007}. Such insulating behavior is due to the the presence of distortion between the TiO$_6$ and CoO$_6$ octahedra as it becomes metallic when the distortion is removed. As we increase pressure, LCTO systematically shows transition from an AFM insulating state to an AFM metallic phase at $\sim$~42~GPa and with further increase in pressure, it becomes a nonmagnetic metal at about 130~GPa. In fact, when the system enters to this metallic state in the AFM phase, a spin-state transition occurs at that particular pressure and hence, Co moment abruptly jumps from its high-spin state value to the low-spin state value. Such insulator to metal transition associated with the spin-state transition is a rare occurrence in AFMs and hence, in case of DP oxides like LCTO, pressure becomes an efficient tool to tailor its physical properties.

The remainder of this paper is written as follows: Sec.~\ref{Experiment_AFM}, we discuss crystal structures and magnetic characterization of the polycrystalline sample in experiment at ambient condition; Sec.~\ref{Methodology}, the methodology for \textit{ab~initio} electronic structure calculations within the density functional theory (DFT) framework is presented; Sec.~\ref{results_ambient_DFT}, we provide the electronic and magnetic properties using fully relaxed atomic coordinates in the unit cell, both for experimental as well as theoretically optimized zero pressure lattice geometries and thereafter, we explode the role of breathing modes in CoO$_6$ and TiO$_6$ octahedron for MIT in the high-symmetric theoretical structure; Sec.~\ref{pressure-effect_structure} finally, within first-principle electronic structure calculations we discuss various details of the structural phase transitions along with the associated magnetic and electronic properties as a function of pressure; at the end, Sec.~\ref{summary} summarizes with conclusion of our work on LCTO.

\section{Experimental results at ambient pressure}
\label{Experiment_AFM} 

\subsection{Crystal structure determination}

A polycrystalline sample of LCTO is synthesized by solid-state reaction route and the structural characterization is performed in the powder sample within X-ray diffraction (XRD) at ambient condition. XRD data are collected using PANalytical Aeris XRD in IIT Tirupati at room temperature, operating Cu $K_{\alpha}$ radiation over the angular range (10$^{\circ}$ $\leq$ 2$\theta$ $\leq$ 90$^{\circ}$) with a $2\theta$ step size of 0.01. Rietveld refinements of the structural parameters were performed using the program FullProf suite software~\cite{Rodriguez-Carvajal1993}. The Rietveld refinement on XRD data of LCTO reveals that it crystallizes in a monoclinic structure with space-group $P2_1/n$ (no.~14) which is consistent with earlier reports~\cite{Rodriguez2002,Holman2007}. The structure consists of corner shared oxygen octahedra forming a three-dimensional (3D) network where Co and Ti are at their center surrounded by six oxygens. On the other hand, La atom is located at the center of the cage formed by the alternating CoO$_6$ and TiO$_6$ octahedra along crystallographic $a, b$ and $c$ directions, see~Fig.~\ref{dist_ambient}(a). It is here evident that in the monoclinic phase, Co-O-Ti chain along $a, b$ and $c$ directions deviates significantly from the ideal/undistorted one by changing the angle, $\angle${Co-O-Ti} from 180$^{\circ}$ to $\sim$ 151--153$^{\circ}$. Table~\ref{tab:structure_exp}~further summarizes all the structural information at room temperature and ambient pressure. The GdFeO$_3$-type distortion is also observed in this sample~\cite{Okeefe1977}. Due to the larger size of Co atom in compared to Ti atom, the CoO$_6$ octahedral volume is expected to be larger than that for TiO$_6$ octahedron.

\begin{table}[!htb]
\centering
\begin{ruledtabular}
\begin{tabular}{cc}
\multicolumn{1}{c}{Parameters} & \multicolumn{1}{c}{Experiment (ambient pressure)} \\  
\hline
Space-group &$P2_1/n$\\
$a$ ({\AA})&5.5565\\
$b$ ({\AA})&5.5717\\
$c$ ({\AA})&7.8566\\
$\beta$ ($^\circ$)&90.002\\
$V$({\AA}$^3$)&243.23\\
$V_\textrm{Co-oct.}$ ({\AA}$^3$)&11.71\\
$V_\textrm{Ti-oct.}$ ({\AA}$^3$)&10.02\\
$V_\textrm{Co-oct.}/V_\textrm{Ti-oct.}$ &1.16\\
$d_\textrm{Co-O}$ (apical)~(\AA)&2.02\\
$d_\textrm{Ti-O}$ (apical)~(\AA)&1.97\\
$d_\textrm{Co-O}$ (planar)~(\AA)&2.08\\
$d_\textrm{Ti-O}$ (planar)~(\AA)&1.98\\
$\angle${\rm Co-O-Ti~(apical)}~($^\circ$)&153.19\\
$\angle${\rm Co-O-Ti~(planar)}~($^\circ$)&154.28\\
\end{tabular}
\end{ruledtabular}
\caption{Table containing structural information (lattice parameters, volumes of the Co- and Ti-octahedra and the ratio between them, bond lengths and bond angles) of the experimental results obtained from the X-ray diffraction at ambient pressure.}
\label{tab:structure_exp}
\end{table}


\subsection{Magnetic measurements}

\begin{figure*}[h!bt]
\centering
\includegraphics[width=2.1\columnwidth]{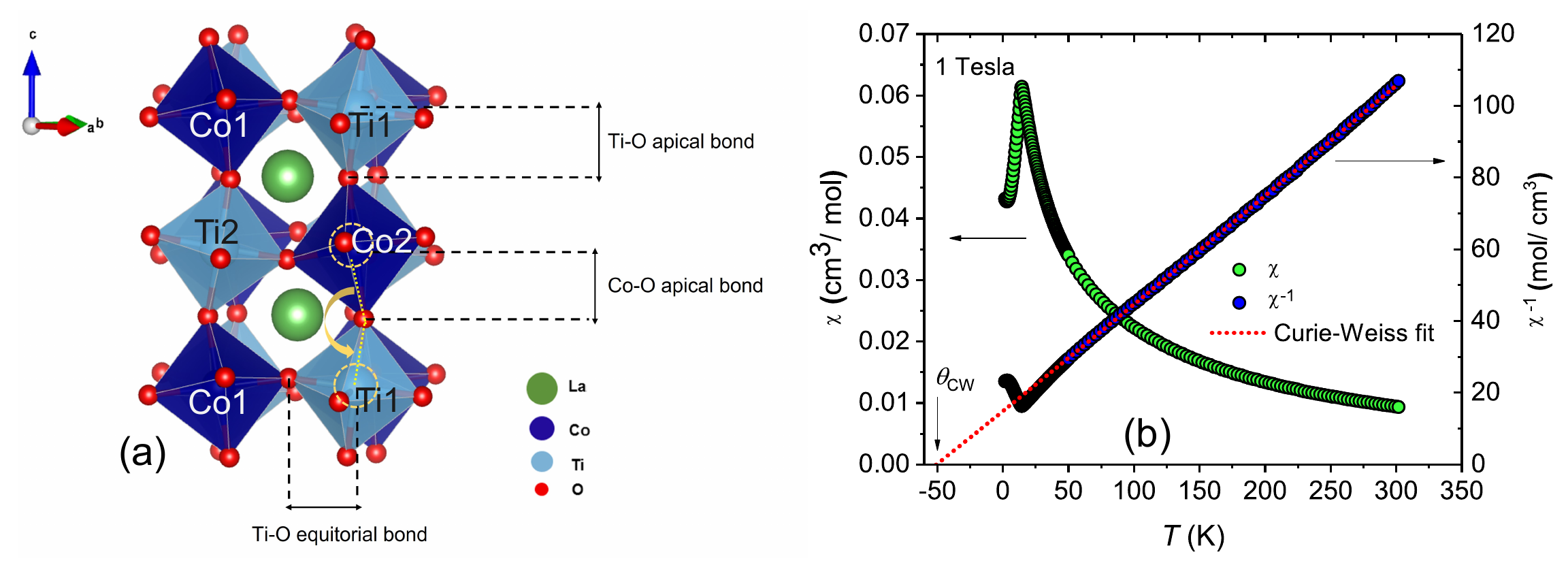}
\caption{\label{dist_ambient} At ambient pressure: (a) structure of LCTO with space-group $P2_1/n$. The angles $\angle${Co-O-Ti} that connect the TiO$_6$ and CoO$_6$ octahedra, both apical and planar, exhibit nearly similar values, approximately 153.2$^{\circ}$ and 154.3$^{\circ}$, respectively. Co and Ti octahedra are distinguished by deep blue and light blue colors, respectively, while La and O atoms are indicated by green and red colors, respectively. (b) The magnetic susceptibility (left axis) and the inverse magnetic susceptibility (right axis) as a function of temperature of LCTO at a magnetic field of $B$ = 1 T. The transition temperature $T_{\mathrm{N}}$ is identified as 14.6 K. Utilizing a Curie-Weiss fit, the Curie-Weiss temperature $\theta_{\mathrm{CW}}$ is determined to be $-$50.5 K, and the effective magnetic moment is estimated at around $\mu_{\mathrm{eff}} \approx 5.05~\mu_{\mathrm{B}}$ within the experimental unit cell of LCTO.}
\end{figure*}

The polycrystalline LCTO is used for the magnetization measurements using a Magnetic Property Measurement System (MPMS3, Quantum Design). Fig.~\ref{dist_ambient}(b) shows the temperature dependent magnetic susceptibility down to 1.8 K in an external field of 1 T. As temperature decreases from the room temperature, the magnetization increases in an usual manner obeying the Curie law in the paramagnetic region. However, the below 14.6 K the magnetic moment drops significantly showing a peak in the magnetization data. This behavior is not only very typical for an antiferromagnetic (AFM) phase transition, but consistent with the literature~\cite{Rodriguez2002,Holman2007,Yuste2011}. Moreover, the Curie-Weiss $\chi = C/(T-\theta_{CW})$, fitting at high temperature between 300 K to 100 K (in the paramagnetic region) yields negative Curie-Weiss temperature, $\theta_{\mathrm{CW}}$ = -50.5 K, confirming AFM interactions between the spins and the obtained effective Co moment, $\mu_{\mathrm{eff}}$ is 5.05 $\mu_{\mathrm{B}}$, will be compared with the value obtained from the DFT (discussed in the next section \ref{results_ambient_DFT}). Detailed experimental findings will be published elsewhere.

\section{Methodology}
\label{Methodology}

The electronic and magnetic properties of LCTO without and with pressure have been explored within \textit{ab~initio} spin-polarized electronic structure calculations using projector augmented wave (PAW) scheme~\cite{Blochl1994,Kresse1999} in the plane wave pseudopotential implementation of DFT code, Vienna \textit{ab~initio} Simulation Package (VASP)~\cite{Kresse1996}. A Perdew-Burke-Ernzerhof (PBE)~\cite{Perdew1996} functional within generalized gradient approximation (GGA) is adopted for the exchange correlation potential. A $k$-mesh of $8 \times 8 \times 6$ ($\Gamma$-centered) is used for the momentum space integration over a full Brillouin zone (BZ), while a plane wave cutoff energy of 500 eV is used for the plane wave basis expansion in all calculations. In order to consider the electron-electron correlation effect on Co-$d$ electrons, GGA+$U$ scheme is employed within Dudarev \textit{et al.} formalism~\cite{Dudarev1998}. Here, we have varied $U$ value between 2 to 6 eV on Co-$d$ ($d^7$) states while on Ti ($d^0$), $U$ is always fixed to zero. The internal position coordinates of all atoms have been fully relaxed until the forces are smaller than 10$^{-3}$ eV/{\AA} and energy convergence cutoff is considered about 10$^{-7}$ eV. In order to determine the optimized zero pressure structure, we perform the total energy calculations of unit cell as a function of its volume. This is performed around the experimental unit cell volume via changing the lattice constants uniformly. According to Birch-Murnaghan (BM) equation~\cite{Birch1947,Murnaghan1937}, a fit to the total energy ($E$) and the unit cell volume ($V$) is valid around the equilibrium geometry. The details of our BM fitting is given in the Appendix~\ref{appendix: BM fit}. The total energy ($E_\mathrm{0}$) corresponding to the zero pressure volume ($V_\mathrm{0}$) point in the BM fitting turns out to be the lowest energy value and the corresponding optimized parameters are further used for the pressure calculations. Achieving true hydrostatic compression requires thorough system relaxation with high precision. Each volume point underwent full geometry optimization until the atomic forces fell below 10$^{-3}$ eV/\AA~for both FM and AFM magnetic configurations within GGA+$U$ calculations. Following structure relaxation, a highly accurate static calculation was conducted using the tetrahedron method with Bl\"ochl corrections and an energy convergence criterion of 10$^{-7}$ eV/ cell. In order to obtain accurate hydrostatic pressure, the structures were fully relaxed with high accuracy. 

\section{DFT results at ambient pressure}
\label{results_ambient_DFT} 

\begin{figure*}[hbt!]
\centering
\includegraphics[width=\textwidth]{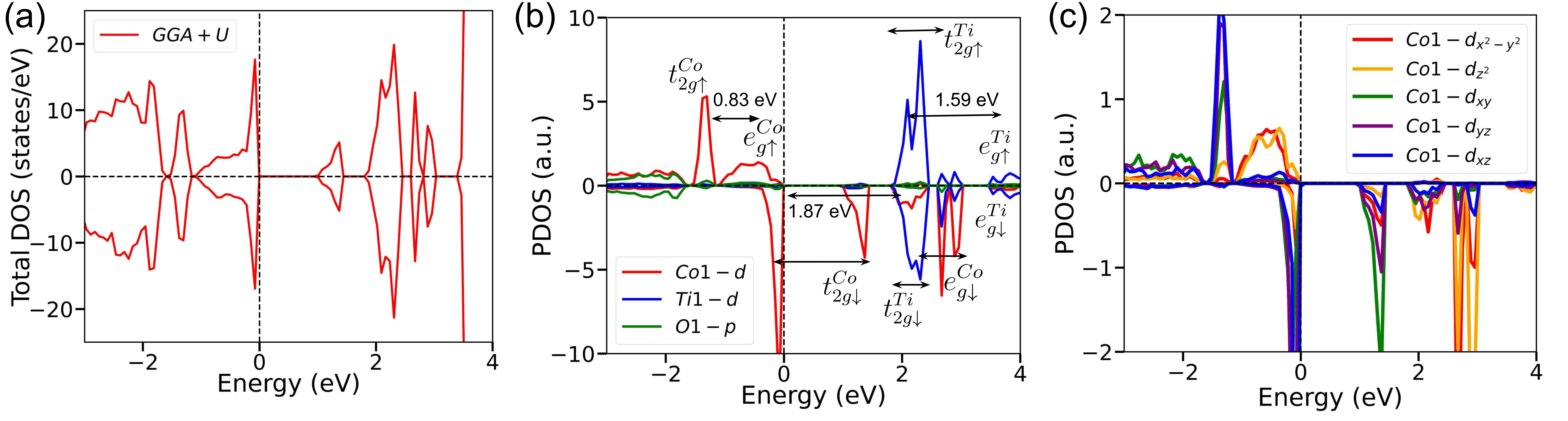}
\caption{\label{dist_ambient_thr} Electronic structures of LCTO considering AFM state using experimental crystal structure at ambient pressure. The total DOSs in (a) exhibit LCTO as insulator within GGA+$U$, $U$ = 2 eV calculations. A gap of 1.01 eV is observed in the calculation with a value of $U$ set to 2 eV. In the context of GGA+$U$ calculations, Fig. (b) depicts the atom-projected partial density of states (PDOS) for the $d$-states of Co1 and Ti1 as well as O-$p$ state. The octahedral crystal field effect on Co and Ti is responsible for the further designation of $d$-state orbitals as $t_{2g}$ and $e_g$. The O-$p$ state is fully occupied, whereas the $t_{2g}$ and $e_g$ orbitals associated with the Ti-$d$ state remain unoccupied. Significantly, the insulating state arises due to the separation of $t_{2g\downarrow}^{Co}$ states, even when considering GGA (not shown) calculations. For GGA+$U$ calculations, (c) also illustrates the orbital-projected DOS for Co-$d$ orbitals. In all cases, the $E_\textrm{F}$ is set at zero.}
\end{figure*}

In order to corroborate with the experimental AFM ground state, in this section, we first discuss the spin polarized electronic structure results calculated using experimental structure (space-group $P2_1/n$) at ambient pressure. The distortion in the experimental structure of LCTO (Fig.~\ref{dist_ambient}(a)) is mainly coming from the octahedral tilting measured by the angle $\angle${Co-O-Ti}, given in table~\ref{tab:structure_exp}. The total DOS in within GGA (not shown) calculation reveals insulting AFM behavior with a gap of 0.21 eV. The magnitude of Co magnetic moment ($\mu_\textrm{Co}$) in the AFM configuration is $\sim$ 2.512 $\mu_\textrm{B}$. With the addition of correlation, $i.e.$, GGA+$U$ calculation for $U$~=~2 eV, gives a slightly enhanced gap value of 1.01 eV (see Fig.~\ref{dist_ambient_thr}(a)) which is indeed in good agreement with the experimental gap value of 1.02 eV~\cite{Holman2007}. By varying $U$, we have further compared both FM and AFM configurations defined within the experimental unit cell as well as the high-symmetric undistorted case, presented in table~\ref{tab:GGA+U_theo_exp}. The total energy of the AFM order is always lower than that of FM state. For $U$~=~2 eV, the Co atom carries an absolute magnetic moment $\mu_\textrm{Co}$ of about 2.6~$\mu_\textrm{B}$.  The Co$^{2+}$ oxidation state under the HS electronic configuration is offering such magnetic moment. This is in agreement with the expected moment per cobalt of 3~$\mu_\textrm{B}$ for the Co$^{2+}$ ($d^7$) high-spin state in the ordered state. The effective magnetic moment $\sim$5.05~$\mu_\textrm{B}$ estimated from the magnetic susceptibility data by fitting the Curie-Weiss law in the paramagnetic region is found to deviate from the spin only contribution (3.87~$\mu_\textrm{B}$) in our spin-polarized calculations. Such discrepancy is well reported for other materials of having Co$^{2+}$ ($d^7$)~\cite{Koo2020,Viola2003,Lee2018}. One thus expects that the system may have some unquenched orbital magnetization which contributes to the total magnetic moment. This is although unlikely to expect significantly large orbital moment in 3$d$ oxides. In order to check the effect of orbital contribution, we have further performed GGA+$U$ calculations with  spin-orbit coupling effect. However, the orbital moment on Co is found to be around 0.17 $\mu_{\mathrm{B}}$ along the z-direction at ambient pressure. Indeed, it is consistent with the fact that the SOC strength for 3$d$ transition metal (here Co)~\cite{Lee2021,Cowley2013} is weak. On the other hand, Ti is found to carry zero magnetic moment, owing to an expected $d^0$ electronic configuration. These are all evident from the atom-projected Co1-$d$, Ti-$d$ and O-$p$ partial density of states (PDOS) presented in Fig.~\ref{dist_ambient_thr}(b) for GGA+$U$ calculations. The $d$-state orbitals of Co and Ti are also marked as $t_{2g}$ and $e_g$ due to the octahedral crystal field effect. Ti-$d$ state is unoccupied whereas the O-$p$ state is fully occupied. These results can be understood by the electronic configuration on Co1 (Co2) site as $t^3_{2g\uparrow}e^2_{g\uparrow}t^2_{2g\downarrow}$ ($t^3_{2g\downarrow}e^2_{g\downarrow}t^2_{2g\uparrow}$), $i.e.$, a partially filled minority spin channel of Co. Moreover, there is a split between the $t_{2g\downarrow}^{Co}$ states, even when considering GGA (not shown) calculations, which is responsible for the insulating behavior of this material. Fig.~\ref{dist_ambient_thr}(c) shows PDOS of Co1 sublattice for different $d$-orbitals.


\begin{figure*}[h!bt]
\centering
\includegraphics[width=\textwidth]{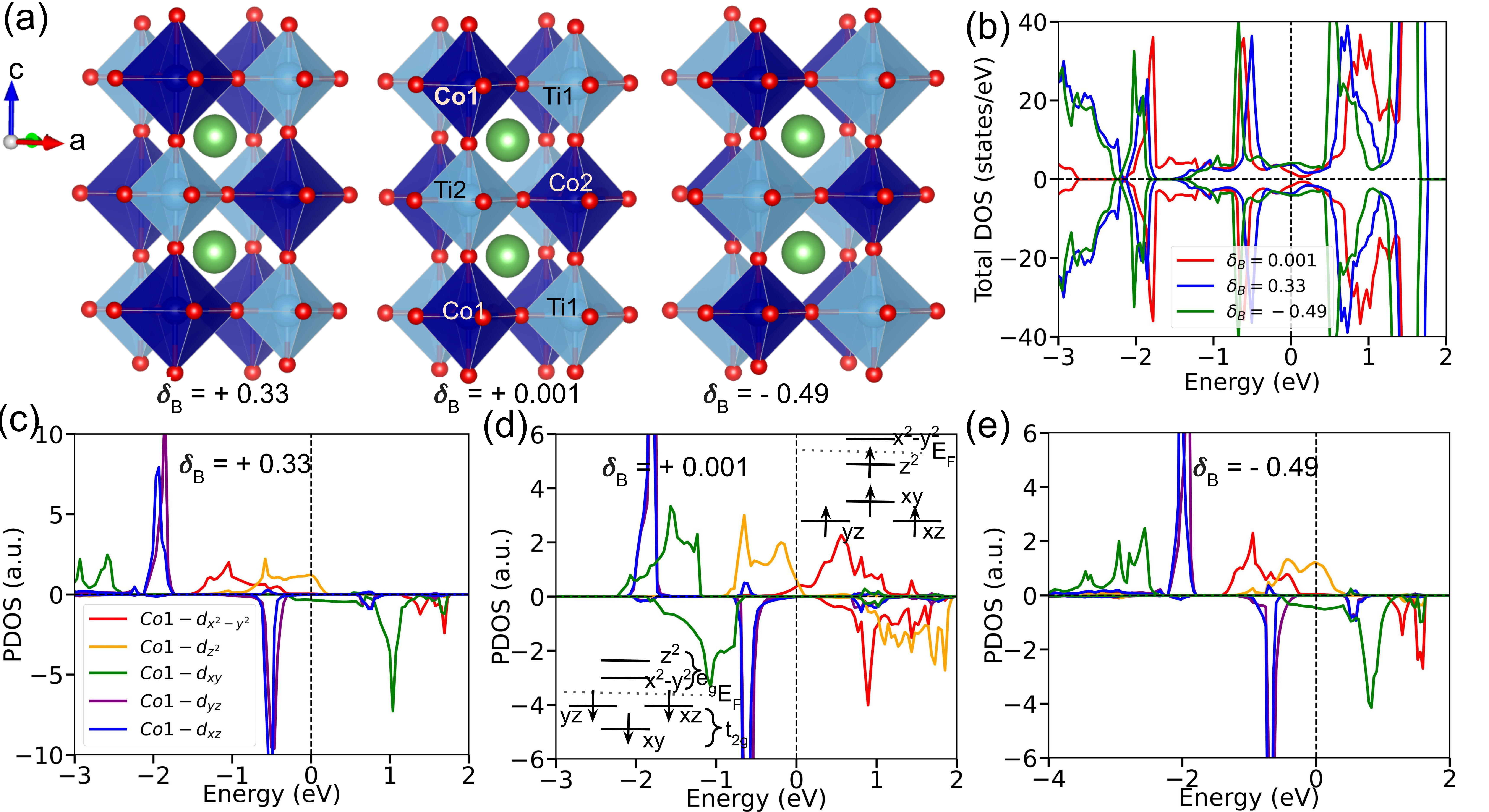}
\caption{\label{undist_ambient} {The impact of a breathing-type distortion on the octahedra of TiO$_6$ and CoO$_6$. In this context, we adopt the high-symmetry $I4/mmm$ structure, leading to an assumed angle of $\angle${Co-O-Ti} = 180$^\circ$. Keeping unit cell volume fixed, the alteration in the volume of the TiO$_6$ and CoO$_6$ octahedra is quantified by introducing the parameter $\delta_\textrm{B}$, defined as $\delta_\textrm{B}=(1 - V_\textrm{Ti}/V_\textrm{Co})$. (a) The configurations displaying the octahedral network of CoO$_6$ (deep blue) and TiO$_6$ (light blue) in LCTO are illustrated for values of $\delta_\textrm{B}$ = $+0.33$, $+0.001$, and $-0.49$. Ti and Co atoms are situated at the center of the octahedra while La and O atoms are marked in green and red colors, respectively. Changing the $\delta_\textrm{B}$ value under the $I4/mmm$ phase does not introduce a gap at $E_\textrm{F}$, as clearly evident from the total DOS shown in (b). However, the orbital-projected DOS considering Co1 in panels (c)-(e) exhibits distinct changes in orbital characteristics as a function of $\delta_\textrm{B}$; for instance, electronic energy levels shift along the energy axis with varying $\delta_\textrm{B}$. We have performed our calculations within GGA+$U$ for $U$ = 2 eV. Remarkably, the $d^7$ electronic configuration of Co remains unchanged despite the presence of the breathing distortion. At $\delta_\textrm{B}$ = 0.001, the magnetic moment on Co is determined to be 0.98 $\mu_\textrm{B}$, supporting a LS state, see the energy level diagram presented in the inset of (d). Within the metallic phase, a significant disparity in octahedral volume ($\delta_\textrm{B} = +0.33$ and $-0.49$) triggers a transition to the HS state, resulting in an increase in the magnetic moment of Co to 0.9 $\mu_\textrm{B}$ and 2.45 $\mu_\textrm{B}$}, respectively. Importantly, throughout all examined cases, the Ti electronic configuration remains consistent at $d^0$ state. The coordinate system in an octahedra is defined such that the position of oxygen atoms align along the x, y and z axes locally, establishing a well-defined reference for orbitals. In that scenario, the $t_{2g}$ manifold comprises with $d_{xy}$, $d_{yz}$, and $d_{xz}$ orbitals, while the $e_g$ manifold comprises with $d_{x^2-y^2}$ and $d_{z^2}$ orbitals.}
\end{figure*}


The structure of the LCTO is distorted, $i.e.$, the angle between $\angle${Co-O-Ti} is $\sim$ 151--153$^{\circ}$, visibly away from the ideal/high-symmetric undistorted structure where the angle between $\angle${Co-O-Ti} is 180$^{\circ}$ (space-group $I4/mmm$). The distortion not only changes the bond angles, also the bond lengths, given in table~\ref{structure_th}. As a matter of fact, the local environments and as well as the interactions are also changed. We studied the breathing mode~\cite{Balachandran2013,Hampel2017} distortion for the undistorted structure by stretching and squeezing the volume of the Co and Ti octahedra periodically. The volume ratio between the Co and Ti octahedra were changed from 5 to 20\% in an equal interval by keeping the unit cell volume fixed. Fig.~\ref{undist_ambient}(a) shows the structures of LCTO for a few selected (maximum breathing in comparison with the no breathing) breathing parameters, defined as $\delta_\textrm{B} = (1 - V_\textrm{Ti}/V_\textrm{Co}$) where the structure remains invariant (space-group $I4/mmm$). Here we discuss the total DOS and orbital-projected PDOS of Co1 for $\delta\rm_B$ = $+0.33$, $+0.001$, and $-0.49$. Fig.~\ref{undist_ambient}(b) shows the total DOS for all the three structures related to the $\delta\rm_B$ = +$0.33$, +$0.001$, and $-0.49$ within GGA+$U$ calculations for $U$ = 2 eV. Interestingly, the undistorted structure shows an antiferromagnetic metallic state, indicating distortion plays an important role to tune the system. The nonzero DOS at $E_\textrm{F}$ confirms that the system is metallic across different $\delta\rm_B$ values. 
Thus, the transition to the metallic state from the insulating state as a function of distortion is accompanied by the electronic phase transition, not driven by the breathing mode distortion, or accompanied by any structural transition. In fact, with the application of hydrostatic pressure, the low-symmetric distorted structure also gives a metallic phase which will be discussed in great details in the subsequent section. Fig.~\ref{undist_ambient}(c), (d) and (e) show the orbital-projected PDOS for Co1 sublattice for $\delta\rm_B$ = $+0.33$, $+0.001$, and $-0.49$ respectively. For $\delta\rm_B$ = 0.001 (see Fig.~\ref{undist_ambient}(d)), $d_{\mathrm{yz}}$ and $d_{\mathrm{xz}}$ orbitals are almost degenerate, however, the down-spin channels are closer to the Fermi level in comparison to the up-spin channels. A schematic orbital energy level diagram is drawn based on the PDOS, indicating a low-spin state (corresponding magnetic moment on Co 0.98 $\mu_\textrm{B}$), shown in the inset of Fig.~\ref{undist_ambient}(d). 
Thus, the spin-state transition, $i.e.$, from a high-spin state (distorted low-symmetric $P2_1/n$ structure) to a low-spin state (undistorted high-symmetric $I4/mmm$ structure) happens with distortion at ambient pressure. We do not expect any orbital order in this system as the Jahn-Teller effect does not exert significant influence on the octahedral geometries around Co1 and Co2, namely, the octahedral geometries are unchanged. Interestingly, there is a significant change in the PDOS as a function of $\delta_\textrm{B}$; $viz.$, electronic energy levels shift along the energy axis with varying $\delta_\textrm{B}$. As a matter of fact, due to the change in the ratios of the octahedral volume induces a high-spin state for $\delta_\textrm{B} = -0.49$ (corresponding magnetic moment in Co 2.45 $\mu_\textrm{B}$) in comparison to the low-spin state for $\delta_\textrm{B} = +0.33$ (corresponding magnetic moment in Co 0.9 $\mu_\textrm{B}$). Importantly, throughout all examined cases, the Ti electronic configuration remains consistent at $d^0$ state. Thus, the breathing mode also plays an important role for transforming the spin-state transition in this material while the structure ($I4/mmm$) remains unchanged. 

\begin{table*}[!htb]
\centering  
\begin{ruledtabular}
\begin{tabular}{ccccccc}
\multicolumn{7}{c}{High-symmetric theoretical structure \textit{vs.} low-symmetric experimental structure comparison}\\ \hline 
\multicolumn{1}{c}{Structure (space-group)} & \multicolumn{3}{c}{Undistorted ($I4/mmm$)} & \multicolumn{3}{c}{Distorted ($P2_1/n$)} \\ 
$U$ value (eV) & 2 & 4 & 6 & 2 & 4 & 6 \\ \hline
$E_{\rm FM-AFM}$ (meV/f.u.) & 270 & 160 & 150 & 190 & 100 & 100 \\
Moment on Co ($\mu_\textrm{Co}$ in $\mu_\textrm{B}$) & 0.98 & 0.99 & 0.99 & 2.60 & 2.71 & 2.80\\
Gap value (eV) & 0 & 0 & 0 & 1.01 eV & 1.58 & 2.08\\
Magnetic ground state & AFM-M & AFM-M & AFM-M & AFM-I & AFM-I & AFM-I \\
\end{tabular}
\end{ruledtabular}
\caption{The competing magnetic states, AFM and FM, are examined as a function of $U$ (with values of 2, 4, and 6 eV) within GGA+$U$ calculations for both the theoretically assumed high-symmetry ($I4/mmm$) structure and the experimentally determined low-symmetry ($P2_1/n$) structure. In both scenarios, the AFM state consistently exhibits lower energy (greater stability) compared to the FM state. It should be noted that the stability of AFM state is examined by considering the most straightforward AFM configuration within a 20-atom unit cell of LCTO. The computed magnetic moment ($\mu_\textrm{Co}$) suggests that the LS (HS) state persists within the $d^7$ electronic configuration of Co for the theoretical structure (the experimental structure) in the electronic configuration of Co. From the gap values in the AFM phase, the last row in the table clearly indicates AFM-M and AFM-I phases for high-symmetric $I4/mmm$ and low-symmetric $P2_1/n$ structures, respectively. The experimentally observed gap is approximately 1.02 eV~\cite{Holman2007}, which corresponds well with the result obtained using $U$ = 2 eV.}
\label{tab:GGA+U_theo_exp}
\end{table*}

The minimum energies obtained for both AFM and FM Co-spins configurations at ambient pressure for both undistorted (theoretically high-symmetric $I4/mmm$) and distorted (experimentally low-symmetric $P2_1/n$) structures within GGA+$U$ calculations for $U$ values of 2, 4, and 6 eV. Table~\ref{tab:GGA+U_theo_exp} consolidates the energy difference between the configurations, magnetic moments, gap values and the corresponding magnetic ground states as a function of $U$ values. For both the structures, AFM remains the ground state providing the lower energy in comparison to FM configuration. The estimated magnetic moment ($\mu_\textrm{Co}$) reveals that the theoretical high-symmetric $I4/mmm$ structure shows a low-spin (LS) state with $\mu_\textrm{Co}$ = 0.98-0.99 $\mu_\textrm{B}$ whereas the experimental low-symmetric $P2_1/n$ structure remains in the high-spin (HS) state with $\mu_\textrm{Co}$ = 2.6-2.8 $\mu_\textrm{B}$. In order to check the effect of orbital contribution in the ordered state, spin-orbit coupling calculations (GGA+$U$+SOC) have been done. However, the orbital contribution from the Co atom is around 0.17 ~$\mu_\textrm{B}$/ Co at ambient pressure. The ground state as a function of $U$ remains at the AFM metallic state for the undistorted $I4/mmm$ structure whereas it is insulating for the distorted $P2_1/n$ structure. In fact, the experimentally obtained gap value 1.02 eV~\cite{Holman2007} corresponding to the gap value obtained for $U$ = 2 eV. As matter of fact, studies under pressure were carried out for $U$ = 2 eV. 

There is a structural transition when the distortion is removed, namely, undistorted tetragonal $I4/mmm$ structure modifies to a distorted cubic $P2_1/n$ structure. However, there is no structural change in the distorted structure when the pressure was applied and the system transforms to a metallic state from an insulating state, will be discussed in the subsequent section \ref{pressure-effect_structure}. There is some similarity with the nickelates system where metal to insulator transition is accompanied by a structural change~\cite{Johnston2014,Hampel2017}. However, undistorted breathing mode structure, $i.e.$, breathing mode in addition to the GdFeO$_3$-type distortion, cannot open a gap at the Fermi level with large $U$ values, cf. table~\ref{tab:GGA+U_theo_exp} in contrast to the nickelates system. Nevertheless, AFM is still remain the ground state of the system.  


\section{DFT results at varying pressure}
\label{pressure-effect_structure} 

\subsection{Structural changes under hydrostatic pressure}

\begin{table*}[h!tb]
\centering 
\begin{tabular}{ccccccccc}
\hline
\hline
\multicolumn{1}{c}{Structural} & \multicolumn{1}{c}{Undistorted} & \multicolumn{5}{c}{Distorted} \\  
parameters & 0 GPa & 0 GPa & 30 GPa & 42 GPa & 60 GPa & 130 GPa \\ \hline
Space-group &$I4/mmm$&$P2_1/n$&$P2_1/n$&$P2_1/n$&$P2_1/n$&$P2_1/n$\\
$a$ ({\AA})&5.3722&5.5621&5.2344&5.1683&5.0812&4.8700\\
$b$ ({\AA})&5.3722&5.5667&5.3411&5.2736&5.1848&4.9007\\
$c$ ({\AA})&7.9659&7.8502&7.4205&7.3254&7.2056&6.9099\\
$\beta$ ($^\circ$)&90.0&90.3&90.3&90.3&90.3&90.3\\
$V$ ({\AA}$^3$)&229.90&243.06&207.65&199.88&189.94&167.24\\
$\kappa$ (\%) &&0&-14.56\%&-17.76\%&-21.85\%&-31.19\%\\
$V_\textrm{Co-oct.}$ ({\AA}$^3$)&9.58&11.81&9.39&8.38&7.78&6.69\\
$V_\textrm{Ti-oct.}$ ({\AA}$^3$)&9.57&10.17&8.98&8.87&8.48&7.38\\
$V_\textrm{Co-oct.}/V_\textrm{Ti-oct.}$ &1.00&1.16&1.04&0.94&0.92&0.90\\
$d_\textrm{Co-O}$ (apical)~(\AA)&2.04&2.06&1.90&1.82&1.79&1.70\\
$d_\textrm{Ti-O}$ (apical)~(\AA)&1.94&1.96&1.88&1.88&1.85&1.76\\
$d_\textrm{Co-O}$ (planar)~(\AA)&1.87&2.07&1.92&1.86&1.80&1.71\\
$d_\textrm{Ti-O}$ (planar)~(\AA)&1.92&1.97&1.89&1.87&1.85&1.77\\
$\angle${Co-O-Ti} (apical)~($^\circ$)&180.00&152.89&156.70&161.79&162.08&169.00\\
$\angle${Co-O-Ti} (planar)~($^\circ$)&180.00&153.51&157.54&162.20&166.32&172.05\\
\hline
\hline
\end{tabular}
\caption{The detailed structural information of LCTO as function of pressure: lattice parameters, unit cell volumes, $\kappa$ = $\frac{(V_{P\neq 0}-V_{P=0})}{V_{P=0}}\times 100\%$, volumes of the CoO$_6$ and TiO$_6$ octahedra, bond lengths and bond angles. Under pressure, $\kappa$ quantifies the percentage change in volume relative to the optimized volume. The undistorted $I4/mmm$ structure is also optimized in our calculation. The pressure values selected (30, 42, 60, and 130 GPa) in the table result in LCTO entering different states or regimes, $e.g.$, AFM-I, AFM-M and NM-M in Fig.~\ref{phase_diag}.}
\label{structure_th}
\end{table*}

\begin{figure*}[!htb]
\centering
\includegraphics[width=2.1\columnwidth]{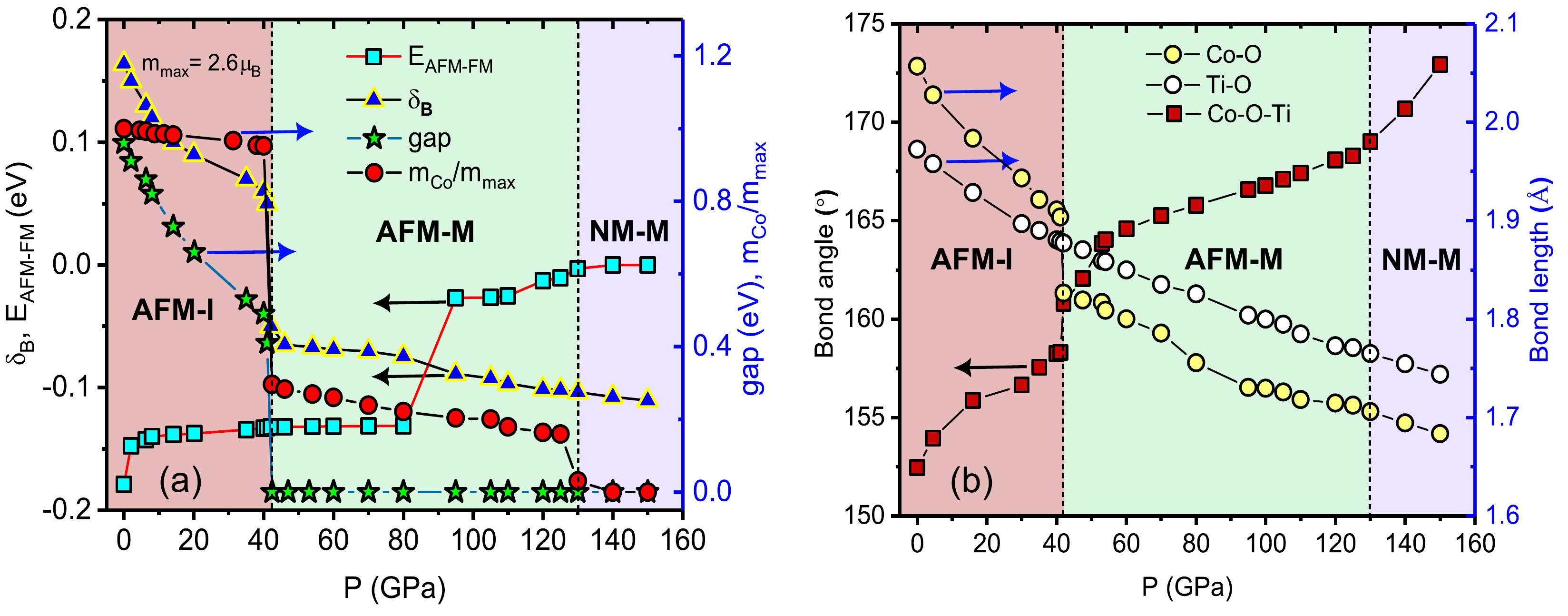}
\caption{\label{phase_diag} Pressure induced variations in various quantities provide support for the existence of various stable phases: AFM insulator (AFM-I), AFM metal (AFM-M) and nonmagnetic metal (NM-M). Three well-defined phases are evident in different colors (light red, light green, and light violet) with their boundary lines clearly depicted as shown in panel (a). The magnetic moment of Co (red circles), expressed in units of m$_\textrm{max}$, experiences a sudden decrease coinciding with a sharp change in $\delta_\textrm{B}$ (blue triangle) around 42 GPa pressure. This transition is accompanied with the gap closing (green star), signifying a transition from an insulator to a metal in the AFM phase. The AFM phase consistently maintains lower energy until the second transition to the NM-M phase occurs around a pressure of 130 GPa, as indicated by the cyan square. Remarkably, a transition from a HS to a LS state in the magnetic moment of Co becomes apparent under pressure, ultimately vanishing in the NM-M phase. In panel (b), we illustrate the pressure-dependent structural changes, specifically along the $c$-axis (apical), including variations in Co-O and Ti-O bond lengths (yellow and white circles, respectively), as well as $\angle${Co-O-Ti} (red square). The simulations are carried out with a $U$ value of 2 eV.}
\end{figure*}

In this section, we discuss the hydrostatic pressure studies, applied maximum 150 GPa on the distorted low-symmetry structure ($P2_1/n$). We tune the system from an antiferromagnetic insulating (AFM-I) state to a nonmagnetic metallic (NM-M) state via an antiferromagnetic metallic (AFM-M) state by the application of hydrostatic pressure. Note that the AFM-M state can also be achieved from the AFM-I state as a function of distortion, discussed in the previous section \ref{results_ambient_DFT}. Table~\ref{structure_th} summarizes all the theoretical optimized structural information, $i.e.$, lattice parameters, unit cell volumes, volumes of the Co- and Ti-octahedra, percentage change in volume defined as $\kappa$ = $\frac{(V_{P\neq 0}-V_{P=0})}{V_{P=0}}\times 100\%$, the ratio between the octahedra volumes, bond lengths (Co-O and Ti-O for both apical and equatorial), and bond angles ($\angle${Co-O-Ti} for both apical and equatorial) used for the calculations. This includes the optimized high-symmetry ($I4/mmm$) undistorted at ambient pressure and low-symmetry distorted ($P2_1/n$) structure at ambient pressure as well as a few selected pressure points (30, 42, 60, and 130 GPa) corresponding to the various regimes, $i. e.$, AFM-I, AFM-M and NM-M of the phase diagram. Under pressure, the monoclinic space-group $P2_1/n$ remains the same with more distorted structure in terms of changes in bond lengths and angles. 

\begin{figure*}[hbt!]
\centering
\includegraphics[width=\textwidth]{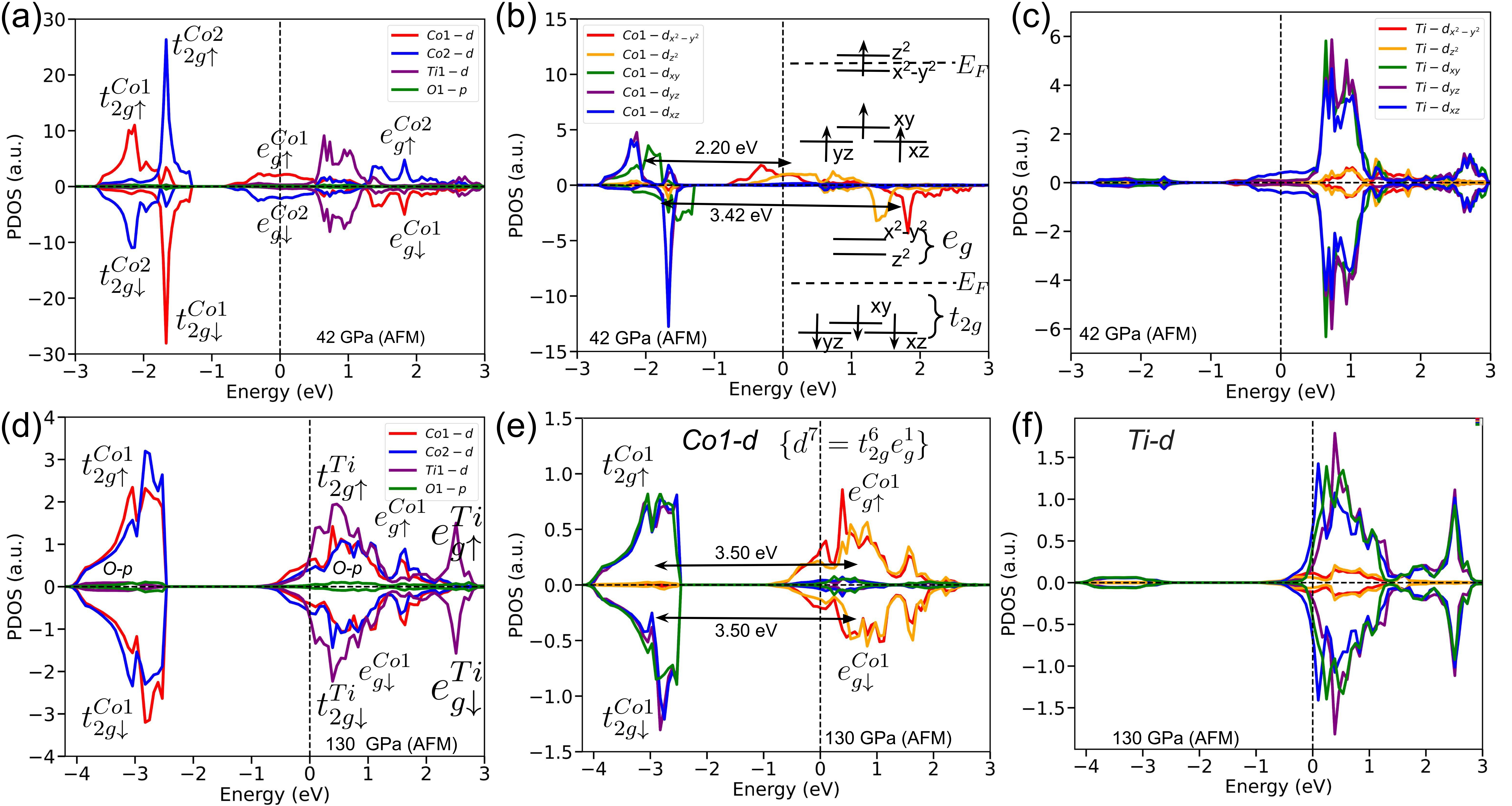}
\caption{\label{pdos_pressure}In the AFM-M phase corresponding to the pressure $P$ = 42 GPa, the following features in the PDOS profile are illustrated: (a) atom-projected PDOS of Co1-$d$, Co2-$d$, Ti-$d$ and O-$p$, (b) orbital-projected PDOS for Co1, (c) orbital-projected PDOS for Ti. The level diagram depicted in the inset of (b), which outlines the orbital nature of the Co-$d$ state, illustrates the LS state of the $d^7$ electronic configuration. Ti remains at $d^0$ state. The development of the magnetic moment within the LS state of Co is linked to the exchange splitting within the $e_g$ level. Nonetheless, this exchange splitting gradually diminishes to zero as pressure continues to rise, typically around 130 GPa. The atom-projected PDOS is presented in (d) for Co1-$d$, Co2-$d$, Ti-$d$, and O-$p$, while the orbital-projected PDOSs for Co1-$d$ and Ti-$d$ are depicted in (e) and (f), respectively. An almost NM-M phase becomes evident at 130 GPa pressure from the calculations, even though a nearly negligible Co moment is observed in our calculations. These calculations are performed within GGA+$U$ for $U$ = 2 eV.}
\end{figure*}

Fig.~\ref{phase_diag} shows a detailed phase diagram containing all the phases, $i.e.$, AFM-I, AFM-M and NM-M as a function of hydrostatic pressure. At ambient pressure, the system is an AFM-I with a band gap of around 1.01 eV at the Fermi level. As pressure increases the band gap decreases almost linearly followed by a sudden jump at around 42 GPa when the system enters to an AFM-M state by closing the band gap, shown in Fig.~\ref{phase_diag}(a). In fact, a number of parameters have been estimated and plotted in the same graph showing significant changes across this phase boundary. The pressure dependent Co moment normalized to its maximum value (2.6 $\mu\rm_B$) decreases linearly till 42 GPa followed by a dramatic change upon entering the AFM-M state. The moment decreases further and at around 130 GPa, the moment abruptly becomes vanishingly small indicating that the system transforms to a NM-M state. The breathing parameter ($\delta\rm_B$) also shows an abrupt jump at the AFM-I to AFM-M phase transition. Interestingly, the jump happens when the volume of the Co and Ti octahedra are almost equal. As the minimum energy for both the AFM and FM configurations is rather small, we calculated the energy difference between the configurations as a function of pressure and plotted in the same graph. Throughout, the AFM phase maintains the lower energy. In fact, there is a slight increase in the energy difference till 130 GPa, then it becomes zero drastically which is consistent with the fact that the system enters to the NM-M state.

Two other parameters, namely, the bond lengths and the bond angles also show interesting consequences when the system enters to this metallic state. Fig.~\ref{phase_diag}(b) shows the pressure variation of the Co-O and Ti-O apical ($c$-axis) bond lengths and $\angle${Co-O-Ti} apical bond angle. Other bond lengths (equatorial bond lengths) and $\angle${Co-O-Ti} equatorial bond angle show (not plotted here for clarity) similar behavior. With the increase of the applied pressure, the Ti-O bond lengths gradually decrease, while the Co-O bond lengths and the $\angle${Co-O-Ti} angle undergo significant abrupt changes at the AFM-I to AFM-M phase transition. In fact, under the application of hydrostatic pressure in a $p$-band oxide Rb$_4$O$_6$, the spin-state transitions ($i.e.$, a transition from a high-spin (HS) state to a low-spin (LS) state and ultimately, to a nonmagnetic state is characterized by the total quenching of the magnetic moment), can develop various magnetic and electronic phases concomitantly: AFM insulator, FM insulator, FM half-metal and finally, nonmagnetic (NM) metal~\cite{Naghavi2012}. Interestingly, in such oxides the magnetism is driven by the $p$-electron-based anionogenic magnetic order where one observes partially filled $\pi$ bands associated with the O$_2$ molecular orbital energy levels~\cite{Nandy2010,Nandy2011,Winterlik2009}. 

Note that all the calculations based on which the phase diagrams are constructed (Fig.~\ref{phase_diag}), were done for $U$ = 2 eV. As electron correlation $U$ plays a vital role for the metal-insulator transition, we also calculated (not shown) the pressure variation of the band gap for different $U$ values. The gap values increase with the increase of the $U$ values and the system takes higher pressure to get into the metallic phase as expected. However, the onset of the antiferromagnetic metallic state is robust for all the $U$ values studied.

At very high pressure ($\sim$ 140 GPa), we tested the structural stability of LCTO through phonon calculations (not shown here) and identified an imaginary optical mode at the $\Gamma$-point. This mode corresponds primarily to oxygen displacement, indicating a pressure-induced ferroelectric-like lattice instability~\cite{Cai2021}. Detailed analysis of pressure-driven ferroelectricity requires extensive study, which could open new avenues for future research in LCTO under pressure.

\subsection{Spin-state transition under pressure and associated electronic structures}
\label{Spin-state-transtion}

As illustrated in Fig.~\ref{phase_diag}, the system exhibits distinct magnetic behaviors under different pressure conditions. To be more specific, the system progresses from an AFM-I phase to an AFM-M phase at $P~\approx$ 42 GPa and subsequently, at $P~\approx$ 130 GPa, it undergoes a further transition, entering a NM-M phase. The initial transition from the AFM-I (as shown in Fig.~\ref{dist_ambient_thr}) to the AFM-M phase is associated with a spin-state transition, during which the Co magnetic moment undergoes a sudden change from its HS state value to the LS state value.
Therefore, we provide a comprehensive examination by analyzing the electronic structures computed for different atoms at two critical pressure values, 42 GPa and 130 GPa. In Fig.~\ref{pdos_pressure}(a), the atom-projected PDOS is particularly depicting the contributions from Co1-$d$, Co2-$d$, Ti-$d$, and O-$p$ states at 42 GPa. Around $E_\textrm{F}$, the Co-atoms, specifically their $d$-orbital, contribute almost entirely compared to the other atoms. Due to the antiferromagnetic alignment of the Co spins in LCTO, the spin polarized $d$-orbital contribution from the Co1 sublattice is exactly opposite to that arising from the Co2 sublattice. As a result, in the AFM-M phase (42 GPa pressure), LCTO exhibits a typical metallic behavior in the total DOS, characterized by the existence of nonzero DOS at $E_\textrm{F}$. Figs.~\ref{pdos_pressure}(b) and (c) present the corresponding orbital-projected PDOS of Co1-$d$ and Ti-$d$ states. As shown in Fig.~\ref{pdos_pressure}(c), it is clear that Ti is in a nonmagnetic state, as its $d$-orbital PDOS features are situated above $E_\textrm{F}$. The electronic configuration of the Ti atom remains as expected $d^0$ configuration, confirming its $4^+$ oxidation state. However, the magnetic moment on Co undergoes an abrupt decrease, reaching a value of approximately 0.97 $\mu_\textrm{B}$. From Fig.~\ref{pdos_pressure}(b), it is crucial to emphasize that the number of $d$ electrons on Co remains unchanged to $d^7$, which in turn leads to a transition in its spin-state due to the applied pressure of about 42 GPa. The corresponding energy level diagram presented in the inset illustrates a LS state for Co1 (Co2) having the electronic configuration as $t^3_{2g\uparrow}t^3_{2g\downarrow}e^1_{g\uparrow}$ ($t^3_{2g\downarrow}t^3_{2g\uparrow}e^1_{g\downarrow}$). Prior to this transition, Co has a magnetic moment of approximately 2.6 $\mu_\textrm{B}$ at pressures below 42 GPa, indicating a HS state with an electronic configuration of $t^3_{2g\uparrow}e^2_{g\uparrow}t^2_{2g\downarrow}$, see also Fig.~\ref{dist_ambient_thr}(b). The reordering of energy levels is correlated with the sudden decrease in the $\delta_\textrm{B}$ value at around 42 GPa pressure, as depicted in Fig.~\ref{phase_diag}(a). Additionally, as pressure is applied, the Ti-O bond lengths gradually decrease, while the Co-O bond lengths and the $\angle${Co-O-Ti} angle undergo significant abrupt changes, as depicted in Fig.~\ref{phase_diag}(b). These changes profoundly affect the electronic structure of LCTO,  particularly the energy level gap between $t_{2g}$ and $e_g$ orbitals on Co. In this context, $t_{2g}$ and $e_g$ levels are ascribed to Co-$d$ based on the splitting occurring within the CoO$_6$ octahedral geometry. In the up-spin channel, the energy separation between $t_{2g}$ and $e_g$ orbitals is initially approximately 0.83 eV under ambient conditions (Fig.~\ref{dist_ambient_thr}(b)) but increases to around 2.2 eV at the pressure corresponding to the transition from AFM-I to AFM-M (Fig.~\ref{pdos_pressure}(b)). A similar trend is also observed in the down-spin channel.
We observe that this results in a crossover of energy levels, with the $e_{g\uparrow}$ level being positioned above the $t_{2g\downarrow}$ level. This, in turn, effectively reduces the exchange splitting within the $t_{2g}$ bands, enabling the emergence of a $t^6_{2g}$ electronic configuration at 42 GPa pressure. Another significant change is observed in the bandwidth of both $t_{2g}$ and $e_g$ levels. The observed bandwidth increase correlates with reduced Co-O bond lengths in the CoO$_6$ octahedra under pressure, enhancing Co-$d$ and O-$p$ state hybridization. However, at ambient condition (HS state), the unequal Co-O (apical) and Co-O (planar) bond lengths within CoO$_6$ octahedra lead to additional splitting of the $t_{2g}$ and $e_g$ levels in the down-spin channel, as seen in the orbital-projected PDOS of cobalt in Fig.~\ref{pdos_pressure}(b). The gap value at $E_\textrm{F}$ within GGA is approximately 0.21 eV, while within GGA+$U$ (see total DOS Fig.~\ref{dist_ambient_thr}(a)), it is approximately 1.01 eV, consistent with its insulating nature. Before transitioning to the LS state with a sudden band gap closure at $E_\textrm{F}$, there is a gradual decrease in the band gap while the system remains in the HS state.
In the LS state, where the Co-$t_{2g}$ level is fully occupied, the extra electron eventually populates the $e_g$ level of Co partially and thereby causing the system to exhibit metallic properties. Therefore, the expected magnetic moment on Co is 1~$\mu_\textrm{B}$ and the calculated moment of 0.97~$\mu_\textrm{B}$ provides strong evidence of the transition to the LS state. In the pressure range approximately from 42 GPa to 130 GPa, LCTO maintains its AFM-M phase with the magnetic moment of Co showing minimal variation, see Fig.~\ref{phase_diag}(a). Note that the energy levels linked to the LS state bears a striking resemblance to the configuration observed in the theoretically optimized undistorted high-symmetry ($I4/mmm$) structure under ambient pressure conditions, see Fig.~\ref{undist_ambient}(d).
 
Once the pressure exceeds 130 GPa, the exchange splitting becomes zero, leading to the complete suppression of the Co moment. This leads to another transition of the AFM-M phase into an NM-M phase at a critical pressure of 130 GPa. 
Furthermore, the phase diagram shown in Fig.~\ref{phase_diag}(a) clearly demonstrates that the energy values for the considered AFM and FM configurations are degenerate. In both cases, the computed Co moment is found to be zero. Fig.~\ref{pdos_pressure}(d) presents the atom-projected PDOS for Co1-$d$, Co2-$d$, Ti-$d$, and O-$p$ states, while Figs.~\ref{pdos_pressure}(e) and (f) depict the corresponding orbital-projected PDOS for Co1-$d$ and Ti-$d$, respectively. Under such high-pressure conditions, it is expected that the hybridization between Co-$d$ and O-$p$ states will be significantly enhanced owing to the further reduction in bond lengths. Comparing Figs.~\ref{pdos_pressure}(a) and (d), a significant expansion in the bandwidths of the $t_{2g}$ and $e_g$ levels can be noted. The Co-$t_{2g}$ orbitals are observed to be pushed even lower with respect to $E_\textrm{F}$, potentially leading to a more significant reduction in the effective exchange splitting. 
Indeed, as demonstrated in the PDOS calculations shown in Fig.~\ref{pdos_pressure}(e), the exchange splitting in the Co-$t_{2g}$/-$e_g$ level has substantially diminished, approaching zero. 
Hence, the magnetic moment on Co is entirely quenched, while the partial occupancy in the $e_g$ band keeps the metallic characteristics in LCTO. The increased bandwidth under pressure can counterbalance the impact of correlation effects ($U$), rendering the absence of a magnetic moment on Co even for a large value of $U$. In other words, the NM-M phase remains stable even with an increase in $U$ up to 6 eV. The results presented here are derived from GGA+$U$ calculations with a $U$ value of 2 eV.

\begin{figure}[h!bt]
\centering
\includegraphics[width=1.0\columnwidth]{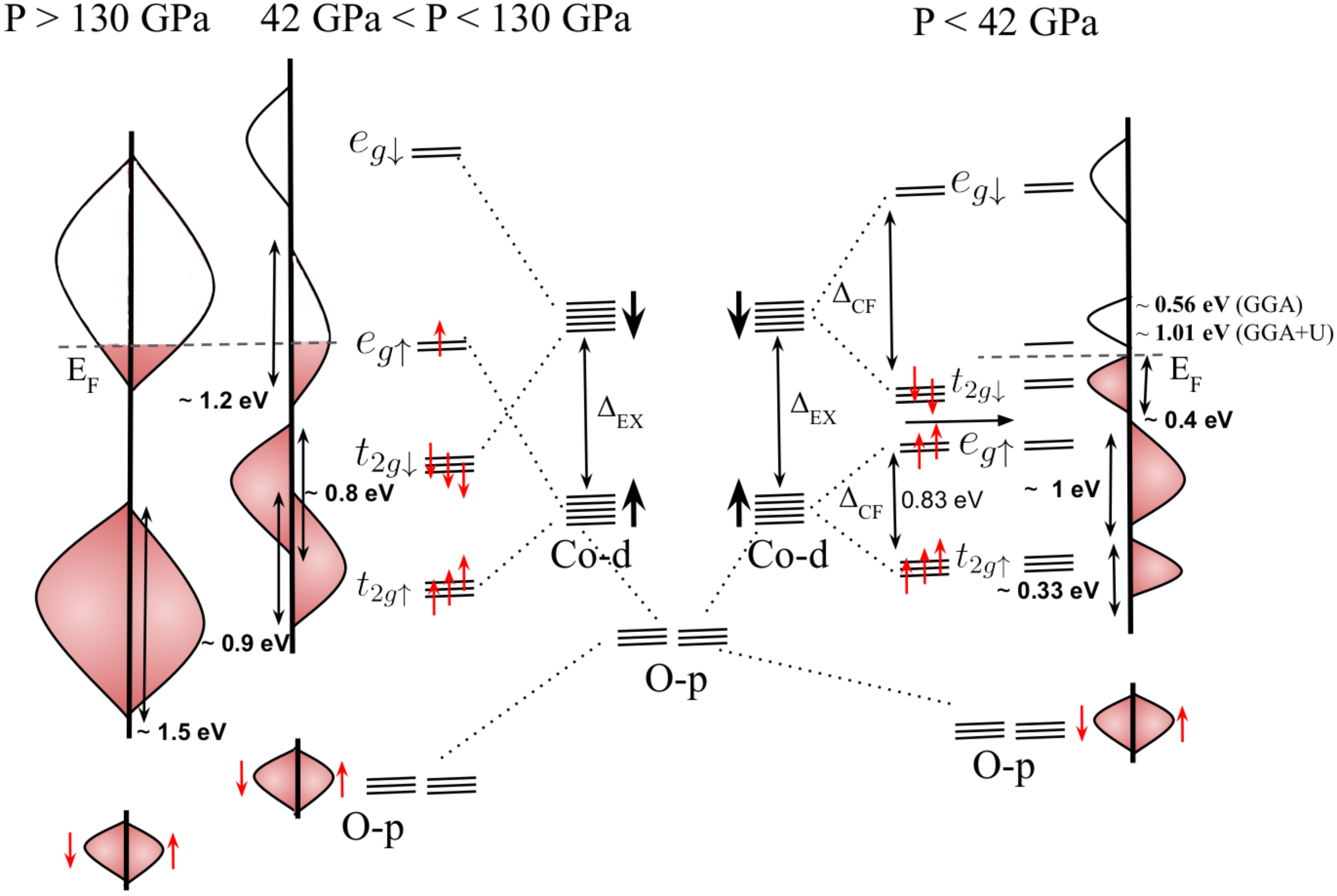}
\caption{\label{fig:LD} The energy level diagram of Co-$d$ orbitals offers insight into the mechanism underpinning the pressure-induced spin-state transition in LCTO, in accordance with the findings from DFT analysis. The energy levels of Co-$t_{2g}$ and -$e_g$ orbitals in both the up and down spin channels are displayed on the right side (HS state), demonstrating the impact of hybridization between Co-$d$ and O-$p$ orbitals within an octahedral arrangement. At the far right, we provide schematics of Co PDOS as computed at ambient pressure. On the left side, the energy levels are adjusted to account for changes in hybridization occurring within a pressure range of 42 to 130 GPa, ultimately resulting in the manifestation of a LS state. Beyond 130 GPa, the magnetic moment of Cobalt (Co) is entirely quenched. The Co PDOSs for 42 and 130 GPa pressures are schematically shown at the far left.}
\end{figure}

Typically, applied pressure leads to reduced inter-atomic distances, causing a decrease in the previously defined breathing parameter, $\delta_\textrm{B}$. As pressure in LCTO increases, the Co-$t_{2g}$ energy level in Figs.~\ref{dist_ambient_thr}(b),~\ref{pdos_pressure}(a), (b), (d) and (e) experiences a discernible shift, as depicted by its position relative to $E_\textrm{F}$. The reduction in CoO$_6$ octahedral volume primarily modifies the octahedral crystal field splitting and simultaneously enhances $d$-$p$ hybridization due to shorter Co-O bond lengths, resulting in increased bandwidth of $t_{2g}$ and $e_g$ orbitals. In Fig.~\ref{fig:LD}, we summarize the pressure dependent studies within an energy level diagram, which illustrates a microscopic mechanism for the spin-state transition. Here, the ultimate energy level diagrams (presented on the right and left sides for HS and LS states, respectively) are illustrated, considering the hybridization between nonmagnetic O-$p$ and Co-$d$ orbitals in the presence of exchange splitting $\Delta_\textrm{EX}$. Furthermore, the final energy levels are designated based on the combined impact of the crystal field and hybridization, with $\Delta_\textrm{CF}$ representing the resulting separation between $t_{2g}$ and $e_g$ levels. The critical aspect of the transition from HS to LS is the energy level crossing between $e^\uparrow_g$ and $t^\downarrow_{2g}$, primarily occurring due to the renormalization in hybridization under pressure. As an illustrative example, the bandwidth of $t_{2g}^\uparrow$ increases nearly threefold, going from around 0.33 eV under ambient conditions to approximately 0.9 eV at 42 GPa, see the schematics of Co PDOS in Fig.~\ref{fig:LD}. The abrupt decrease in $\delta_\textrm{B}$ at about 42 GPa pressure (see Fig.~\ref{phase_diag}) signifies that concurrent with the crossover, a metallic behavior in LCTO appears with the $E_\textrm{F}$ lies in the $e^\uparrow_g$ state. The $e^\uparrow_g$ state exhibits a notably wide bandwidth of approximately 1.6 eV, indicating the robustness of the metallic state even when we increase the electron correlation parameter $U$ to 6 eV. At the AFM-M to NM-M transition point (130 GPa pressure), the $t_{2g}$ level is pushed even further below $E_\textrm{F}$, with a $\Delta_\textrm{CF}$ of approximately 3.50 eV, as illustrated in Fig.~\ref{pdos_pressure}(e). The bandwidth, which is also approximately 1.5 eV, is a consequence of even stronger hybridization effects. In this situation, our calculations reveal complete quenching of the exchange splitting in the $t_{2g}$ and $e_g$ levels, maintaining the electronic configuration as $t^6_{2g}e^1_g$ (LS state), but resulting in a zero magnetic moment on the Co site. The mechanism depicted in Fig.~\ref{fig:LD} hence demonstrates how changes in pressure significantly impact the energy levels and bandwidths of Co-$t_{2g}$ and $e_g$ orbitals, leading to a series of transitions: from AFM-I to AFM-M, and ultimately to NM-M. In addition, we have estimated the Hund coupling (J) and crystal field splitting ($\Delta$) values from the PDOSs in three different pressure regions: 30 GPa (AFM-I), 42 GPa (AFM-M), 130 GPa (NM-M), tabulated in table \ref{tab:table4}. This estimation is carried out by analyzing roughly the center of bands, $t_{2g}$ and $e_g$ in the PDOS.  The separation, E($t_{2g\uparrow}^{Co1}$) - E($t_{2g\downarrow}^{Co1}$), for two different spin channels provides an approximate estimation of J, while the separation, E($t_{2g\uparrow}^{Co1}$) - E($e_{g\uparrow}^{Co1}$), within the same spin channel offers an estimate of the crystal field splitting $\Delta$~\cite{Saha-Dasgupta2020,Sandratskii2003,Sahakyan2022}. The values as a function of pressure exhibit a systematic variation, clarifying the spin state transition in LCTO. In future, we would like to do a very detailed study on this aspect by performing model calculations.
 
\begin{table}[!htb]
\begin{ruledtabular}
\begin{tabular}{ccd}
\multicolumn{1}{c}{\textrm{P (GPa)}}&
\multicolumn{1}{c}{\textrm{Estimated J value (eV)}}&
\multicolumn{1}{c}{\textrm{Estimated $\Delta$ value (eV)}}\\
\hline
8&1.16&0.83\\
42&0.6&2.2\\
130&0.0&3.5\\
\end{tabular}
\end{ruledtabular}
\caption{The variation of Hunds coupling J and crystal field splitting $\Delta$ in three different regions, namely, at 8 GPa (AFM-I), 42 GPa (AFM-M), 130 GPa (NM-M).}
\label{tab:table4}
\end{table}

\section{Summary and conclusions}
\label{summary} 

The double perovskite oxide La$_2$CoTiO$_6$ was first examined experimentally through synthesis and magnetic measurements, followed by a detailed investigation using first-principle electronic structure calculations within density functional theory. This comprehensive study aimed to understand the magnetic and transport properties of La$_2$CoTiO$_6$ under different external pressure conditions. At ambient pressure, the structure is monoclinic with space-group $P2_1/n$ and it shows an insulating behavior with antiferromagnetic ground state below $T_\textrm{N}$ = 14.6 K at ambient pressure. However, with the application of hydrostatic pressure on the experimental structure, studied within first-principle electronic structure calculations, we can tune the magnetic and transport properties, $viz.$, antiferromagnetic insulator  to antiferromagnetic metal  at around 42 GPa, and finally to itinerant nonmagnetic metal at around 130 GPa. Interestingly, antiferromagnetic insulator  to antiferromagnetic metal transition is accompanied with a spin-state transition (high-spin to low-spin state) while Co and Ti remain in the $d^7$ and $d^0$ electronic configuration, respectively. 
The spin-state transition under pressure is found to be robust with $U$ up to 6 eV. The structural distortion, particularly a breathing mode distortion near the critical pressure point, facilitates the spin-state transition as the system transitions into the metallic phase. Furthermore, we observe a low-spin state even under ambient pressure conditions in a theoretically high-symmetric structure where the lattice distortions in the monoclinic structure are removed. The introduction of breathing mode by changing the ratio of the Co and Ti-octahedra governs an increase in Co magnetic moment in the metallic phase. From the PDOS and the corresponding energy level diagram illustrate that the pressure-induced spin-state transition  is a result of both the octahedral crystal field splitting in Co-$d$ orbitals (splitting to $t_{2g}$ and -$e_g$ levels) and the bandwidths, mainly driven by the strong hybridization between Co-$d$ and O-$p$ orbitals. The nonmagnetic metallic phase at high pressure (above 130 GPa) is characterized by the total quenching of the Co magnetic moment. Thus, La$_2$CoTiO$_6$ is a rare antiferromagnetic double perovskite oxide where the insulator to metal transition can be tuned as a function of hydrostatic pressure/distortion/even changing the breathing parameter which is accompanied by a spin-state transition. In the future, experimental validation of these phase transitions can be conducted by subjecting the synthesized material to hydrostatic pressure conditions.

\begin{acknowledgements}

R.S.M. acknowledges the financial support from SERB for his Early Career Research Award (File no.: ECR/2018/000999/PMS). The authors would like to acknowledge IIT Tirupati for providing experimental setup for the sample preparation, high performance computing (HPC) cluster for DFT calculations and IIT Mandi for the magnetization measurements. In addition, HPC support from IISER Tirupati is also acknowledged. A.K.N. and S.P. acknowledge the support from National Institute of Science Education and Research (NISER), Department of Atomic Energy, Government of India, for funding the research work through project number RIN-4001. A.K.N. and S.P. acknowledge the computational resources, Kalinga cluster, at NISER, Bhubaneswar, India. A.K.N. thanks Prof. P. M. Oppeneer for the Swedish National Infrastructure for Computing (SNIC) facility and VASP simulations.

\end{acknowledgements}

\appendix

\section{Birch-Murnaghan (BM) fitting}
\label{appendix: BM fit} 

As pressure increases, the unit cell volume decreases, leading to a reduction in the bond lengths between atoms. The total energy obtained from self-consistent calculations for specific volumes was fitted to the Birch-Murnaghan (BM) isothermal equation of state~\cite{Birch1947,Murnaghan1937}, as shown below. 

\counterwithin{figure}{section}
\begin{figure}[h!bt]
\centering
\includegraphics[width=1.15\columnwidth]{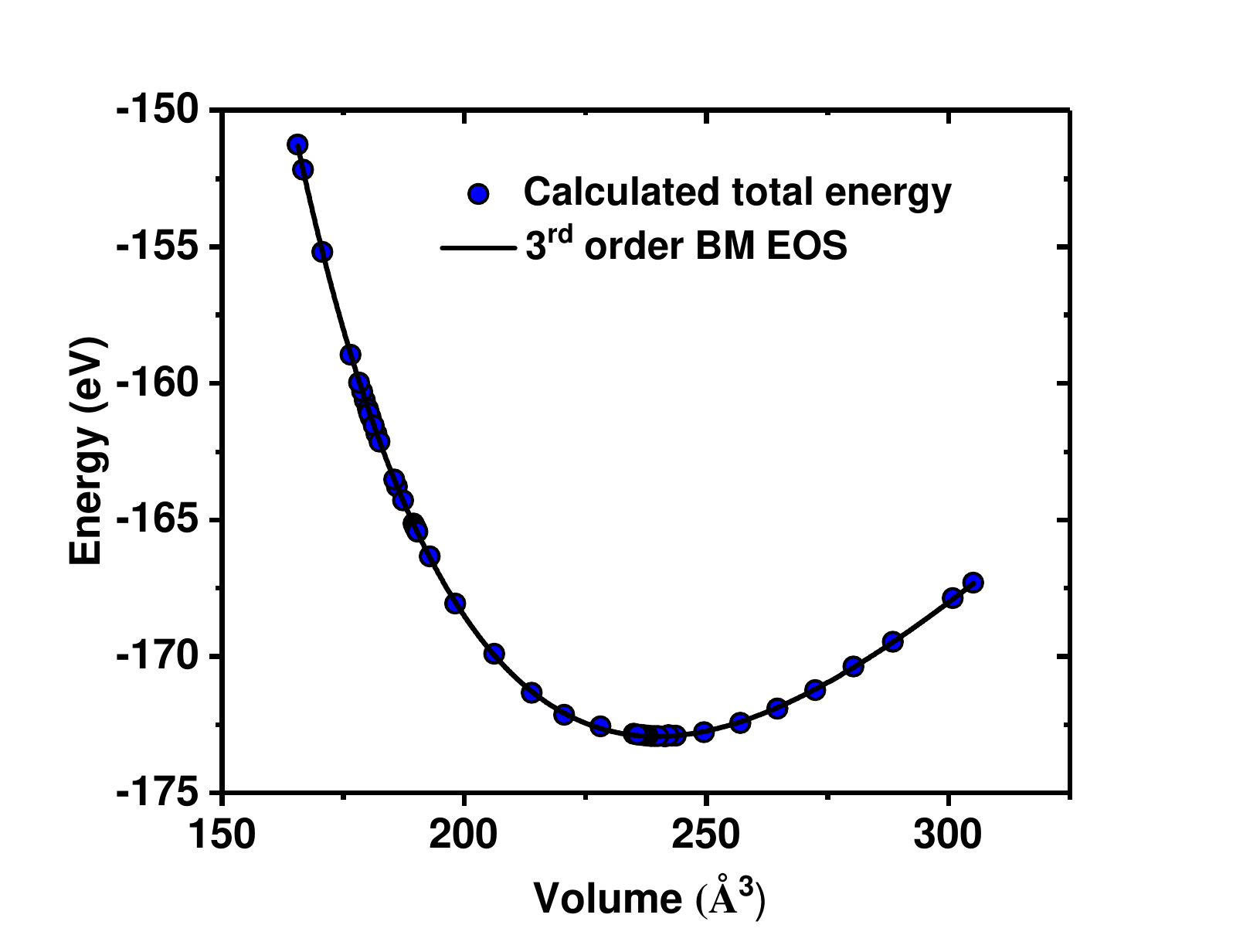}
\caption{\label{fig:BM_fit.pdf} The \textit{ab~initio} calculated energy ($E$) versus volume ($V$) data of La$_2$CoTiO$_6$ within GGA+$U$ calculation considering $U$ = 2 eV. This is fitted to the Birch-Murnaghan equation.} 
\end{figure} 

\begin{eqnarray}
E(V) = & E_0+\frac{9V_0B_0}{16}\left[\left(\frac{V_0}{V}\right)^\frac{2}{3}-1\right]^3 B_0^{\prime}\\ \nonumber
& + \!\left[\left(\frac{V_0}{V}\right)^\frac{2}{3}-1\right]^2 \left[6-4\left(\frac{V_0}{V}\right)^\frac{2}{3}\right],
\end{eqnarray}
where $B$ and $B^{\prime}_\mathrm{0}$ are the bulk modulus and its pressure derivative, respectively and $E_\mathrm{0}$ is the total energy at the equilibrium volume $V_\mathrm{0}$ \cite{Birch1978,Beekman2022,Zhang2023}.

By fitting the \textit{ab~intio} computed energy value as a function of volume, the calculated parameters are as follows: $E_\mathrm{0}$ = -172.92 eV, $V_\mathrm{0}$ = 239.87 \AA$^3$, $B_\mathrm{0}$ = 142.78 GPa, and $B^{\prime}_\mathrm{0}$ = 4.42.

To determine the pressure, the derived fitting parameters were then incorporated into the pressure-volume BM relation~\cite{Birch1947,Murnaghan1937} as,

\begin{eqnarray}
P(V) = & \frac{3B_0}{2}\left[\left(\frac{V_0}{V}\right)^{7/3}-\left(\frac{V_0}{V}\right)^{5/3}\right]\\ \nonumber
& \times \!\biggl\{1+\frac{3}{4}\left(B_0^{\prime}-4\right)\left[\left(\frac{V_0}{V}\right)^{2/3}-1\right]\biggl\}.
\end{eqnarray}
We have performed these calculations within GGA+$U$ for $U$ = 2 eV where the structure remains unaffected with space-group $P2_1/n$. 

The ambient pressure lattice parameters for La$_2$CoTiO$_6$, which represent the ground state parameters, are determined through the BM fitting shown in Fig.~\ref{fig:BM_fit.pdf}. The smallest volume point on the plot corresponds to the La$_2$CoTiO$_6$ structure at zero pressure, and volumes less than this correspond to the structure under pressure.


\end{document}